\documentclass[aps,,twocolumn,floatfix,amsmath,amssymb,showpacs]{revtex4}
\usepackage{txfonts}
\usepackage{amssymb}
\usepackage{graphicx}
\usepackage{dcolumn}
\usepackage{bm}
\usepackage{color}

\newcommand{\Rmnum}[1]{\expandafter\@slowromancap\romannumeral #1@}
\newcommand{\etal}{\emph{et al.}}
\newcommand{\be}{\begin{equation}}
\newcommand{\ee}{\end{equation}}
\newcommand{\bfig}{\begin{figure}}
\newcommand{\efig}{\end{figure}}

\begin{document}

\title{Effect of hydrostatic pressure on transport in the topological insulator Bi$_2$Te$_2$Se}

\author{Yongkang Luo$^{1,2}$, Stephen Rowley$^{1,*}$, Jun Xiong$^{1}$, 
Shuang Jia$^{3}$, R. J. Cava$^{3}$, and N. P. Ong$^{1}$
}

\affiliation{
$^{1}$Department of Physics, Princeton University, Princeton, New Jersey 08544, U.S.A,\\
$^{2}$Department of Physics, Zhejiang University, Hangzhou 310027, P. R. China, \\
$^{3}$Department of Chemistry, Princeton University, Princeton, New
Jersey 08544, U.S.A
}

\begin{abstract}
The Hall coefficient $R_H$ and resistivity $\rho$ of the topological insulator Bi$_2$Te$_2$Se  
display a number of puzzling features vs. temperature $T$. We propose a model that describes well
the non-monotonic variation of $R_H(T)$. A key feature of the model is a chemical potential
that is weakly $T$-dependent. From the fit to the model, we 
infer a ``transport'' gap $\Delta_T$ of 50 mV. We find that hydrostatic pressure
$P$ (0-27 kbar) has a pronounced effect on both $R_H$ and $\rho$. We show that
these changes arise from decreases in both $\Delta_T$ and the hole
effective mass under pressure. 
\end{abstract}


\date{\today}

\maketitle

Topological Insulators are a class of semiconductors in which the
bulk energy gap is traversed by current-carrying surface states, which have 
a massless, Dirac dispersion. A novel feature of the surface state is 
the locking of the spin of the surface electron perpendicular to its momentum 
with a helicity that has opposite signs on opposing faces of a crystal
~\cite{Bernevig-Science06,Fu-PRL07,Fu-PRB07,Moore-07,Fu-PRL08,Qi-PRB08}.
The spin locking strongly suppresses back-scattering. In several Bi-based semiconductors,
the existence of the topological surface state (SS) and its spin-locked nature
have been established by angle-resolved
photoemission spectroscopy
(ARPES)~\cite{Hsieh-BiSb,Hsieh-Sb,Xia-Bi2Se3,Chen-Bi2Te3}. Detailed 
scanning tunneling microscopy (STM) experiments have also confirmed the
spin-locked nature of the surface states~\cite{Roushan-STM}. 
In transport experiments, Shubnikov-de Haas (SdH) oscillations from 
the SS have been detected in 
Bi$_2$Te$_3$~\cite{Qu10}, and in (Bi,Sb)$_2$Se$_3$~\cite{Analytis}. 
Analysis of the SdH oscillations in Bi$_2$Te$_3$~\cite{Qu10} 
yields a surface mobility $\mu_s$ = 8,000-10,000 cm$^2$/Vs.
Despite the high mobility, the bulk conductance $G_b$ can exceed 
the surface conductance $G_s$ by factors of 3000 or more in crystals of these materials.
A major experimental task is to bring the ratio $G_b/G_s$ significantly below 1.

\begin{figure}[t]
\includegraphics[width=10cm]{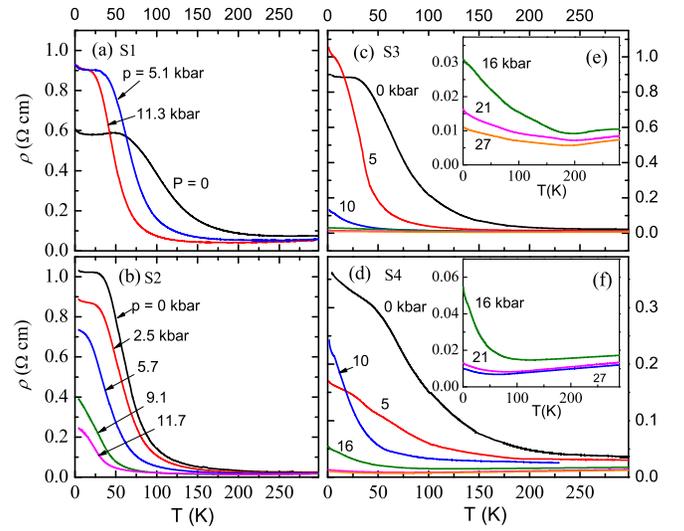}
\caption{\label{figrho}
(Color online) The resistivity profiles, $\rho$ vs. $T$, at selected pressures 
in Bi$_2$Te$_2$Se for samples S1$\cdots$S4 (Panels a$\cdots$d, respectively). 
Resistivities of Samples S3 and S4 at higher $P$ are shown in Panels (e) and (f), 
respectively. In sample S1, the curves were 
measured in the sequence $P$ = 0, 11.3 and 5.1 kbar. In all panels,
$\rho$ is plotted along the vertical axes, and $T$ along the horizontal axes.
}
\end{figure}

Recently, the hybrid semiconductor Bi$_2$Te$_2$Se has emerged as a 
highly promising topological insulator (TI) for transport experiments~\cite{Ando10,Xiong1}. 
ARPES measurements~\cite{Suyang} reveal a single spin-locked Dirac state crossing a bulk
energy gap of $\sim$350 mV. The Dirac point lies close to the top of the 
bulk valence band at the $\Gamma$ point.
In carefully annealed crystals, the 
observed resistivity $\rho$ at 4 K (Fig. \ref{figrho}) is typically 1 $\Omega$cm, but can be
much larger. In Refs. \cite{Xiong1,Xiong2}, we show that 
in crystals with $\rho$(4 K) = 6 $\Omega$cm, the surface SdH oscillations are
especially well-resolved.
The surface SdH oscillations yield a surface mobility $\mu_s$ = 2,800-3,200 cm$^2$/Vs, 
which implies that the ratio $G_b/G_s$ is $\sim$1. 
At 4 K, the residual bulk carriers are $n$-type with a 
bulk density $n_b$ equal to $\sim 3\times 10^{16}$ cm$^{-3}$
(compared with 10$^{18}$--10$^{19}$ cm$^{-3}$ in Bi$_2$Te$_3$ and Bi$_2$Se$_3$).
Hall measurements show that the bulk mobility $\mu_b$ is
60 times smaller than $\mu_s$ at 4 K~\cite{Ando10,Xiong1}.

\begin{figure}[t]
\includegraphics[width=9.5cm]{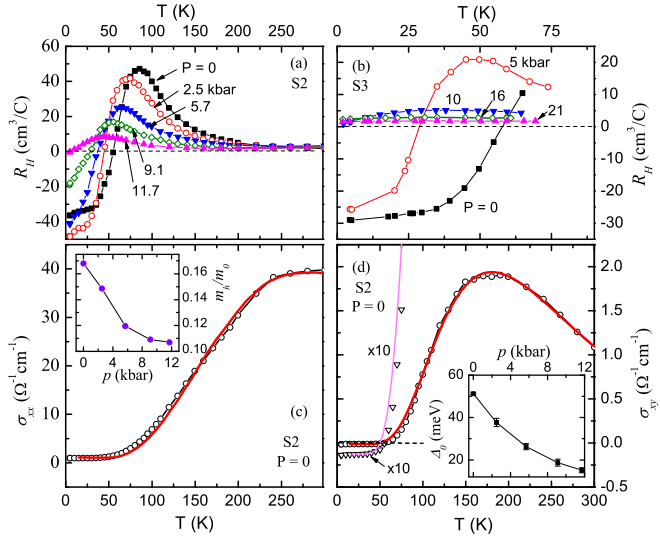}
\caption{\label{figHall}
(Color online) The temperature dependence of the Hall
coefficient $R_H$ in samples S2 (Panel a) and S3 (Panel b) at the pressures indicated.
The solid curves in (a) and (b) are guides to the eye.
Panels (c) and (d) plot the conductivity $\sigma_{xx}$ and Hall conductivity 
$\sigma_{xy}$ derived from $R_H$ and $\rho$ in S2 at $P$ = 0 (open symbols). The bold
curves are the fits of the data to the proposed model. In Panel (d), we replot for clarity
the low-$T$ data (triangles) and the fit (thin curve) in expanded scale ($\times$10).
The effective mass $m_h$ and transport gap $\Delta_T(0) = \Delta_0$ derived from the fits
in S2 are plotted vs. $P$ in the insets in Panels (c) and (d), respectively. 
}
\end{figure}

Despite the large $\rho$ in Bi$_2$Te$_2$Se, much of its bulk conduction features 
are puzzling. The Hall coefficient $R_H$ displays an
interesting non-monotonic profile vs. $T$ that seems unrelated to thermal activation at first glance
(Fig. \ref{figHall}). To further improve the quality of Bi$_2$Te$_2$Se crystals, a better 
understanding of where the bulk carriers come from is clearly important. 
We propose a model that accounts well for the $T$ dependence of the weak-field conductivity
tensor. We show that the transport quantities are consistent with thermal activation across a small transport 
gap of $\sim$50 mV. A key feature of the model is a weakly $T$-dependent chemical potential $\mu(T)$.
Hydrostatic pressure allows us to tune the system parameters. We show that the
observed changes are consistent with the model. In addition, we uncover an interesting
trend under pressure in the magnetoresistance.

Four crystals of Bi$_2$Te$_2$Se, grown by a
modified Bridgman method~\cite{Ando10,Xiong1} (Samples S1$\cdots$ S4), were investigated. 
Two piston-cylinder pressure cells were used, with high-purity 
Pb or Sn serving as the manometer. The pressure was determined
by the pressure-dependent superconducting transition temperature of Pb (or Sn), as well as the resistance ratio
$R(p)/R(0)$ at room temperature. Thin crystals with mirror-like surfaces 
(nominally 1.5 $\times$ 0.6 $\times$ 0.02 mm in size) were 
cleaved from the ingot. The cleaved faces expose the outer Te layers. The axes $\bf a$ and
$\bf b$ lie in the layer, and the axis $\bf c$ is normal to the cleavage plane.  Ohmic contacts were
made in the Hall-bar configuration on each crystal.
With daphne oil as the pressure fluid, we attained a maximum hydrostatic pressure
of 27 kbar.

Figure \ref{figrho} plots the temperature dependence of $\rho$ under
pressure. At ambient pressure ($P$ = 0), the observed resistivity 
$\rho$ initially increases steeply as the temperature $T$ is decreased from 300 K, but tends towards
a plateau value below 50 K. This saturation 
implies a parallel metallic conduction channel that has been identified to be
the topological surface states~\cite{Ando10,Xiong1}.
With applied pressure, the $\rho$-$T$ profile generally becomes much more metallic 
at all $T$. We observe two classes of behavior. In S2, $\rho$ at 4 K decreases monotonically
with increasing $P$ (Fig. \ref{figrho}b), whereas, in S1, S3 and S4, the change is
non-monotonic if $P<$ 10 kbar (Panels a, c and d). However, above 10 kbar, $\rho(T)$
tends towards a metallic profile with a weak $T$ dependence in all samples.

Pressure also has a pronounced effect on the Hall coefficient $R_H$, as shown in Fig. \ref{figHall}a
(in S2) and Fig. \ref{figHall}b (S3). At $P$ = 0, the curve $R_H$ vs. $T$ in S2
displays a characteristic peak at 80 K and a zero-crossing at 58 K~\cite{Ando10}. 
In applied pressure, both features shift monotonically to lower $T$. Simultaneously,
the large negative plateau at low-$T$ moves upwards, eventually attaining a positive
value at 11.7 kbar. For S3, $R_H$ is already positive at all $T$ at 10 kbar.
To exploit the additivity of the conductivities, we invert the resistivity tensor $\rho_{ij}$
to obtain the conductivity tensor $\sigma_{ij}$. The inferred conductivity $\sigma_{xx}$ and Hall conductivity
$\sigma_{xy}$ are plotted as open circles in Panels (c) and (d), respectively.
Also plotted are the fits (bold curves) to these quantities using the following model.

As may be seen in the expanded scale in Panel (d), $\sigma_{xy}$ (open triangles)
is negative and nearly $T$-independent below 30 K. Above 30 K, 
$\sigma_{xy}$ undergoes an exponential increase to very large positive values. This suggests
that we have a small density of $n$-type carriers at low $T$, which 
are overwhelmed by a divergent population of holes induced by thermal activation 
across a small gap. However, because $\sigma_{xy}$ rises to a broad peak at 180 K,
the exponential growth is eventually tempered by high-$T$ processes.

\begin{figure}
\includegraphics[width=7.5cm]{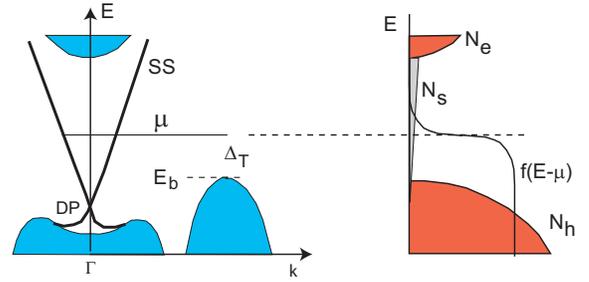}
\caption{\label{figband}
(Color online) Sketch of the band structure near the $\Gamma$ point. The left panel shows 
the Dirac surface state (SS) as bold lines.
The chemical potential $\mu$ lies $\sim$130 mV above the Dirac point (DP).
Shaded regions represent bulk bands. 
The maximum of the valence band lies below $\mu$ by the 
energy spacing $\Delta_T$ (the maximum is likely at a finite $\bf k$). 
The fits show that $\mu$ is $T$ dependent. The right panel shows the
densities of states of the bulk conduction and hole bands (${\cal N}_e$ and ${\cal N}_h$, 
respectively) and the surface states ${\cal N}_s$. 
The solid curve labelled as $f(E-\mu)$ is the Fermi-Dirac distribution. 
}
\end{figure}

\begin{table}[h]
\caption{\label{tab:table1} Fit parameters at selected pressures $P$ (for Sample S2).
$\Delta_0$ is the transport gap at $T$ = 0, $\gamma$ the coefficient of the $T^2$
term in the gap, $m_h$ the hole effective mass (with $m_0$ the free electron mass)
and $\beta$ is the exponent in the hole mobility.}
\begin{center}
\begin{ruledtabular}
\begin{tabular}{ccccc}
       $P$ (kbar)  & $\Delta_0$ (meV) & $\gamma$ (meV/K$^2$) & $m_{h}/m_0$ & $\beta$  \\ \hline
       0           &     51         &   6.30$\times 10^{-4}$  &  0.168   &     1.79        \\
       2.5         &     38         &   7.70$\times 10^{-4}$  &  0.149   &     0.86      \\
       5.7         &     26         &   7.52$\times 10^{-4}$    &  0.119   &     0.71      \\
       9.1         &     19         &   7.17$\times 10^{-4}$    &  0.109   &     0.76     \\
       11.7        &     15         &   5.76$\times 10^{-4}$    &  0.107   &     1.30  \\
\end{tabular}
\end{ruledtabular}
\end{center}
\end{table}

At 4 K, we assume that charge conduction involves carriers in the
topological surface state and a dilute concentration of bulk carriers
(both $n$-type). As found in Ref. \cite{Xiong1},
the bulk carriers, which have very low mobility ($\mu_b\sim$ 50 cm$^2$/Vs), occupy an impurity band.
By contrast, the high mobility of the surface electrons ($\mu_s\sim$ 3,000 cm$^2$/Vs)
leads to SdH oscillations (see below). The data at 4 K, restricted to relatively low $B$, 
do not allow the surface and bulk electrons to be separated (in Refs. \cite{Qu10,Xiong1}, the
separation was accomplished by using the SdH oscillation amplitudes), 
we treat the two bands here as one band with an effective mobility $\mu_e = A/(1 + C T^{\alpha})$, with
$\alpha$, $A$ and $C$ as free parameters.

The conductivity is $\sigma_{xx} = 
2e^2\sum_{\bf k}(-\partial f/\partial E) v_x({\bf k})^2\tau$, 
where $f(E-\mu)$ is the Fermi-Dirac distribution, ${\bf v(k)}$ the group velocity
at the wavevector $\bf k$ and $\tau$ is the transport lifetime.
With the chemical potential $\mu$ in the bulk gap as sketched in Fig. \ref{figband}, 
we obtain for the holes
$\sigma_{xx}  = n_h e\mu_h$ and $\sigma_{xy} = n_h e\mu_h^2 B$, with $\mu_h$ the hole
mobility.
In the limit $\Delta_T/k_BT\gg 1$ ($k_B$ is the Boltzmann constant) 
the hole population has the activated form
\be
n_h(T) = \frac{1}{4} \left(\frac{2 m_h k_B T}{\pi \hbar ^2}\right)^{\frac32}
{\rm e}^{-\Delta_T/k_B T},
\label{nh}
\ee
where $m_h$ is the hole effective mass. The energy scale $\Delta_T \equiv \mu-E_b$ 
is the ``transport gap'' that dictates the exponential growth in $\sigma_{xy}$ at low $T$
($E_b$ is the energy at the valence band maximum). As we show below, we may 
ignore thermal activation to the conduction band for $T<$ 300 K. 
We express the hole mobility as $\mu_h = DT^{-\beta}$,
with $D$ and $\beta$ as free parameters (the impurity scattering rate for the holes plays
no role in the fitting because the holes are thermally activated).

To fit the data at high $T$, however, we further assume that $\Delta_T$ has the 
temperature dependence
\be
\Delta_T(T) = \Delta_0 + \gamma T^2,
\label{DT}
\ee
with $\Delta_0$ and $\gamma$ as free parameters.

The fits to the conductivity tensor are plotted as bold curves in Figs. \ref{figHall}c and \ref{figHall}d.
We remark that our primary goal is to describe the weak-field hole conductivity tensor
using the 4 constants $\Delta_0$, $\gamma$, $m_h$ and $\beta$.
The strong thermal activation at low $T$ and the pronounced non-monotonicity 
of $\sigma_{xy}$ at high $T$ severely restrict the physically reasonable values that these 
parameters may assume. Hence the convergence of the fits is rapid and relatively unambiguous
(by contrast, the $n$-band parameters $\alpha$, $A$ and $C$ are less reliably obtained
because the $n$-type carriers are resolved only below 30 K). We have extended the fits to finite
$P$ as well for S2. The optimal parameters are reported in Table I.

From the fits, we infer that the steep exponential increase in $\sigma_{xy}$ when $T$ is
raised above 30 K (triangles in Fig. \ref{figHall}d) results from the thermal activation
of holes into the valence band (Eq. \ref{nh}). The monotonic decrease of the hole
mobility $\mu_h$ with increasing
$T$ tends to counter this increase, especially in $\sigma_{xy}\sim \mu_h^2$.
However, to produce the non-monotonic variation in $\sigma_{xy}$, we need $\Delta_T$
to increase slightly with $T$ as in Eq. \ref{DT}. We attribute this 
increase to a $T$-dependent chemical potential. Particle number conservation 
dictates that $\mu$ must change with $T$
when $k_BT$ is large enough to reach a band with a large density of states (here
the hole band). This causes $\mu$ to move away from the valence band with an increment 
varying as $T^2$. At ambient $P$, the increment is 6.3 mV when $T$ = 100 K (Table I).
As a consequence, $n_h(T)$ grows much more slowly above 200 K, resulting in an overall decrease
in $\sigma_{xy}$ as observed. We expect a $T$-dependent
$\mu$ to be a general feature in topological insulators because $\mu$ lies close to one of the
bulk band extrema.

\begin{figure}[t]
\includegraphics[width=9.5cm]{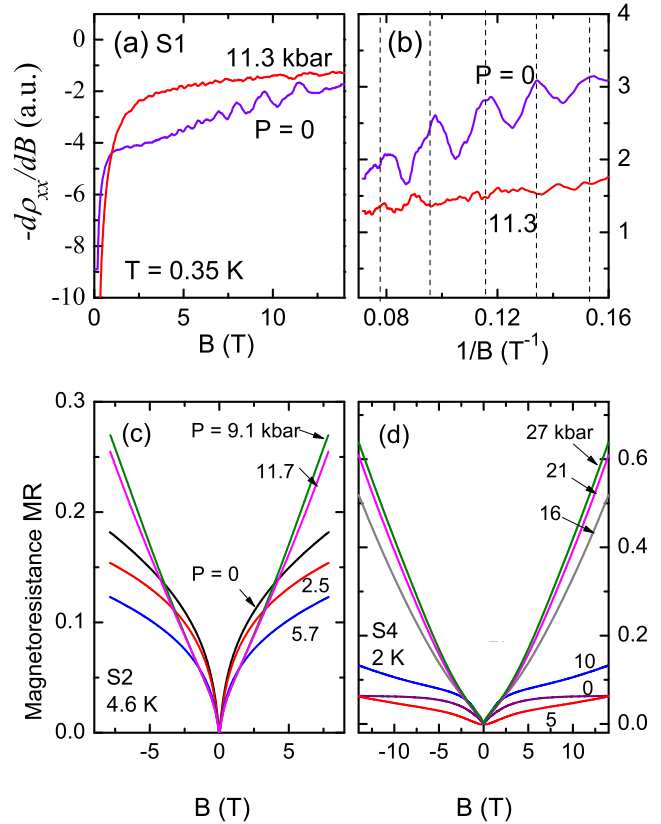}
\caption{\label{figSdH}
(Color online) Comparison of the magnetoresistance curves at 
ambient pressure and high pressure. In Panel (a), the derivative $d\rho_{xx}/dB$ at $P$ = 0
and $T$ = 0.35 K displays SdH oscillations from the surface states. Panel (b) shows that the 
oscillations are periodic in $1/B$, with a period that yields $k_F$ = 0.040 \AA$^{-1}$.
The oscillations become unresolved at 11.7 kbar. 
In Panels (c) and (d), we plot the MR curves (measured with $\bf B||c$) at selected pressures for Samples
S2 (at 4.6 K) and S4 (2 K), respectively. In both samples, the MR 
changes from a saturation trend (at low $P$) to one that is large and quadratic in 
$B$ at large $P$.
}
\end{figure}

As in Refs. \cite{Qu10,Ando10,Xiong1}, we may estimate the position of $\mu$ by 
detecting the weak surface SdH oscillations (Fig. \ref{figSdH}a). From the period
of the oscillations at 0.35 K, we obtain the surface Fermi wavevector $k_F$ = 0.040 \AA$^{-1}$.
Using the surface Fermi velocity $v_F$ = 6$\times 10^5$ m/s~\cite{Xiong1,Suyang}, 
the $k_F$ value implies that $\mu$ lies 130 mV above the Dirac point. ARPES~\cite{Suyang} shows that
the direct gap at the $\Gamma$ point ($\bf k=0$) is 350 mV. Hence
the conduction band minimum lies 220 mV above $\mu$. As this separation exceeds
$\Delta_T$ by a factor of $\sim$4, we are justified in ignoring thermal 
excitations into the conduction band. In addition, we note that
$\Delta_T$ is quite small compared to the spacing of $\mu$ from the Dirac point (130 mV).
This implies that the valence-band 
maximum is at a finite $\bf k$, as drawn in Fig. \ref{figband}.

From the fits, we now see that the dramatic effect of $P$ on the transport properties
stems primarily from the steep decrease of the transport gap. In Sample S2 (inset in Fig. \ref{figHall}d), 
the zero-Kelvin value $\Delta_0$ decreases from 51 mV (at $P$ = 0) to 15 mV ($P$ = 11.7 kbar). 
Thus, at 11.7 kbar, the smallness of $\Delta_0$ results in a large 
population of excited holes even at 4 K ($R_H$ is close to zero at 11.7 kbar).
Nonetheless, a vestige of the activated behavior can be seen at 11.7 kbar in $\rho$ vs. $T$
(Fig. \ref{figrho}c). At larger $P$ (16-27 kbar),
$\Delta_T$ is nearly completely suppressed (Fig. \ref{figrho}d, for S4), and the
system remains $p$ type down to 4 K. In addition, the fits reveal that an increasing $P$ decreases the 
hole effective mass from 0.16 times the free mass (at $P$ = 0) to 0.11 (at 11.7 kbar). 
The most significant effect, however, is that increasing $P$ suppresses the surface conductance, rendering 
it difficult to resolve against the increased hole conduction. Direct evidence for this
suppression is obtained by comparing the curves of $d\rho_{xx}/dB$ 
(Fig. \ref{figSdH}a,b). The prominent SdH oscillations at $P$ = 0
become unresolved at 11.3 kbar. Thus, in closing the transport gap, pressure
converts the system from a TI in which surface conductance $G_s$ is easily observed at 0.35 K to
one that is dominated by the bulk holes.

Tuning the relative weights of $G_s$ and $G_b$ by pressure provides
a powerful way to separate transport characteristics of the surface states from bulk states. 
As an example, we examine the transverse magnetoresistance, defined
as the fractional change in $\rho(B)$, viz.
$MR(B) = [\rho(B)/\rho(0)]-1$, measured with $\bf B||c$. Figures \ref{figSdH}c and \ref{figSdH}d plot 
the curves of $MR(B)$ at several pressures for Samples S2 (at 4.6 K) 
and S4 (at 2 K), respectively. In both samples, when $G_s$ and $G_b$ are comparable (at low $P$), 
the MR tends towards saturation at large fields. However, when $G_b$ dominates (large $P$),
the MR switches to a quadratic variation ($\sim B^2$) consistent with the semiclassical
MR observed for bulk states in a high-mobility semi-metal. In both samples, 
the dominance of the bulk conduction at large $P$ is again evident in the 
large, semi-classical $B^2$ variation of $\rho$. 
Currently, there is very little understanding of the MR of both the bulk
and surface states, as discussed in Ref.~\cite{Qu10}.
The present results suggest that the behavior of the MR with pressure tuning may 
reveal specific characteristics of the topological surface state, and 
provide a way to readily distinguish surface from bulk conduction.

Finally, we remark that the fits provide a clearer picture of the overall 
band structure in crystals of Bi$_2$Te$_2$Se that have $\mu$ inside the 
bulk gap. We obtain an accurate determination of the gap $\Delta_T(T)$, which
dictates the bulk carrier concentration above 30 K, as well as its sensitivity to
applied pressure. It should be instructive to see how these numbers are altered when the pure
compound is doped with donors or acceptors. An interesting test of the inferred
band alignments may be made with thermopower, which is 
sensitive to the band parameters.

We acknowledge support from the National Science Foundation under
Grant DMR 0819860 and from the Nano Electronics Research Corporation
(Award 2010- NE-2010G). Y.K.L. acknowledges a
scholarship given by the China Scholarship Council (CSC).\\

\noindent
\emph{$^*$Current affiliations of SR:}
Centre for Materials and Microsystems, Fondazione Bruno
Kessler, I-38100 Trento, Italia, and 
Cavendish Laboratory, University of Cambridge, United Kingdom.

\end{document}